# Positron studies of surfaces, structure and electronic properties of nanocrystals

**S.W.H. Eijt**[*,1], **B. Barbiellini**[2], **A.J. Houtepen**[3], **D. Vanmaekelbergh**[3], **P.E. Mijnarends**[1,2], and **A. Bansil**[2]

[1] Faculty of Applied Sciences, Delft University of Technology, Delft, Netherlands
[2] Department of Physics, Northeastern University, Boston, MA 02115, USA
[3] Institute of Physics and Chemistry of Nanomaterials and Interfaces, Utrecht University, Utrecht, Netherlands

A brief review is given of recent positron studies of metal and semiconductor nanocrystals. The prospects offered by positron annihilation as a sensitive method to access nanocrystal (NC) properties are described and compared with other experimental methods. The tunability of the electronic structure of nanocrystals underlies their great potential for application in many areas. Owing to their large surface-to-volume ratio, the surfaces and interfaces of NCs play a crucial role in determining their properties. Here we focus on positron 2D angular correlation of annihilation radiation (2D-ACAR) and (two-detector) Doppler studies for investigating surfaces and electronic properties of CdSe NCs.

**1 Introduction** Nanoscale particles have become extremely important in the tailoring of materials for a wide range of applications [1-3]. In particular, semiconductor nanocrystals (NCs) have attracted wide interest because their electronic and optical properties can be tuned by variation of their size and shape [2, 3]. At small sizes, the surface or interface energy becomes a very significant factor in establishing not only the interior crystal structure but also the specific surface structure and ad-atom termination of the NCs. These special properties underlie the large potential of NCs for applications in areas of nanostructured thin-film solar cells [4], opto-electronics [3] and spintronics [5], fission/fusion reactor vessel steels [6], metal hydrides for hydrogen storage [7], and fluorescence-based detection of biomolecules [3].

During the last decade, it has been shown that positrons ($e^+$) act as a sensitive, self-seeking probe of embedded and colloidal NCs [8-20], revealing the local structural properties and electronic structure via a measurement of the electron momentum density [21]. Positron methods are thus capable of extracting valuable information on properties at specific surface, interface or defect sites in the NCs, which are difficult to access by other methods such as transmission electron microscopy, X-ray diffraction (XRD), and EXAFS. Furthermore, insight into the electronic structure of buried or embedded NC systems is gained, which is difficult to obtain with e.g. STM or XPS.

**2 Positrons as a sensitive probe of nanocrystals** The large sensitivity with which positrons probe embedded NCs derives from their diffusive motion by which they scan the host matrix and finally trap in one of the NCs before annihilation. In positron beam studies [22], positrons are implanted at a high ini-tial kinetic energy –typically in the range of a few keV– but they rapidly loose their energy through interactions with electrons and phonons and thermalize in a time short compared to their average lifetime in the material. Subsequently, they may trap in one of the embedded NCs if the difference in the positron affinity $A_+ = -(\phi_+ + \phi_-)$ [22, 23] of NC and host material is negative and sufficiently large. Here, $\phi_+$ and $\phi_-$ are the $e^+$ and $e^-$ work functions, respectively, defined for the same zero energy level in the crystal [23].

___________
[*] Corresponding author: e-mail: s.w.h.eijt@tudelft.nl, Phone: +31 15 278 9053, Fax: +31 15 278 6422

The case of diffusion-limited trapping serves to illustrate the high sensitivity in positron studies of NCs. The observed fraction of positrons $f_{NC}$ annihilating with electrons associated with the NC is given in this model by [10, 22]:

$$f_{NC} = \frac{\kappa}{\kappa + \lambda_{host}} = \frac{4\pi r_{NC} D_+ c_{NC}}{4\pi r_{NC} D_+ c_{NC} + \lambda_{host}} = \frac{4\pi r_{NC} L_+^2 c_{NC}}{4\pi r_{NC} L_+^2 c_{NC} + 1} = \frac{3(L_+/r_{NC})^2 f_V}{3(L_+/r_{NC})^2 f_V + 1} \quad (1)$$

with $\kappa$ is the impurity trapping rate and $\lambda_{host}$ is the annihilation rate in the host matrix. $\kappa$ is furthermore specified in terms of the $e^+$ host diffusion constant $D_+$, the NC radius $r_{NC}$ and the concentration of NCs $c_{NC}$. Finally, $f_{NC}$ can be expressed in the positron diffusion length $L_+ = \sqrt{D_+ \tau}$, with $\tau = \lambda_{host}^{-1}$ the $e^+$ lifetime in the host matrix, $r_{NC}$ and the specific NC volume fraction $f_V$. Clearly, the positron signal $f_{NC}$ stemming from the NCs is significantly amplified relative to the NC fraction $f_V$ for $e^+$ diffusion lengths $L_+$ large compared to $r_{NC}$. Positron studies on embedded Cu NCs in Fe [8] and Li NCs in MgO [9, 10] indeed show observed fractions of positrons annihilating in NCs >90%, while $f_V$ is only a few %. Consequently, the positron signal carries predominantly information on the NCs rather than on the matrix. It is therefore important to ascertain whether specific embedded NC systems show positron trapping. In this connection, *ab initio* methods can be employed to calculate the positron affinities $A_+$ [22-24,11,10] of NC and host materials. Such computations have been performed for a wide variety of bulk materials [22, 23].

Figure 1 shows schematically four examples of affinity profiles: (a) bcc-Cu NC embedded in Fe [8]; (b) fcc-Li NC embedded in MgO [9, 10]; (c) Au NC embedded in MgO [15-17]; and (d) colloidal CdSe NC [11, 12, 14]. Calculated bulk $e^+$ affinities from Refs. [10, 11, 21-23] were used. Affinity values for vacancy clusters at the boundaries have not been calculated and therefore serve only as an illustration. Nagai *et al.*, using the two-detector Doppler and 2D-ACAR methods, observed nearly complete confinement of the positron inside ~1 nm Cu NC embedded in an Fe matrix [8] (Fig. 1(a)). These studies showed that interface defects such as vacancies were absent because of the near-coherent fit of bcc-Cu NCs in the Fe host matrix (~1% lattice mismatch). Similarly, positrons are confined in ~2-5 nm fcc-Li NCs created in MgO single crystals by means of ion implantation and subsequent thermal treatment [10, 20] (Fig. 1(b)). Despite a good fit to the host matrix, 2D-ACAR studies strongly indicate the presence of vacancy defects at the Li/MgO interface [9] as additional trapping centres. The latter is also observed for Au NCs in MgO [15-17], where the presence of vacancies in fact enables positron methods to reveal the otherwise hidden properties of the Au/MgO interface (Fig. 1(c)). Since the $e^+$ affinity value of Au is somewhat higher than that of MgO, trapping in the Au NCs themselves is not expected. Positron lifetime and (two-detector) Doppler depth-profiling studies by Xu *et al.* [15, 16] identified the trapping centres as $V_4$ vacancy clusters residing at the Au/MgO interface. 2D-ACAR and two-detector Doppler studies on high-temperature annealed samples [17] also excluded trapping inside Au NCs, and showed that vacancy trapping only occurs at the interfaces of Au NCs and not in neighbouring annealed MgO layers.

Finally, Fig. 1(d) shows the example of colloidal CdSe NCs, which themselves may act as a strong positron trap. However, it was recently demonstrated that the surface acts as a strong positron trapping center and the positron density in the interior of the NC system therefore remains limited [14]. The positron remains in a shell-like surface state [14], and can thus be used as a new probe to detect surface composition and modifications in colloidal NC systems. Electron confinement effects can also be clearly observed [11] since the tails of the confined Se(4p) wave functions extend into the surface shell layer where they are detected because of overlap with the positron wave function at the surface. In this latter case, diffusion is clearly unimportant considering the large concentration of NCs stacked in the films. Also, the sizes of these NCs are typically an order of magnitude less than positron diffusion lengths of ~50-100 nm in inorganic semiconductors. Therefore, trapping is transition-limited [22].

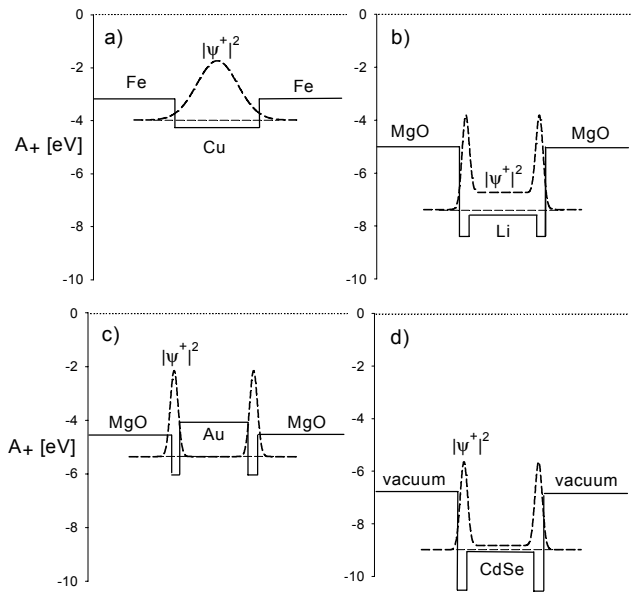 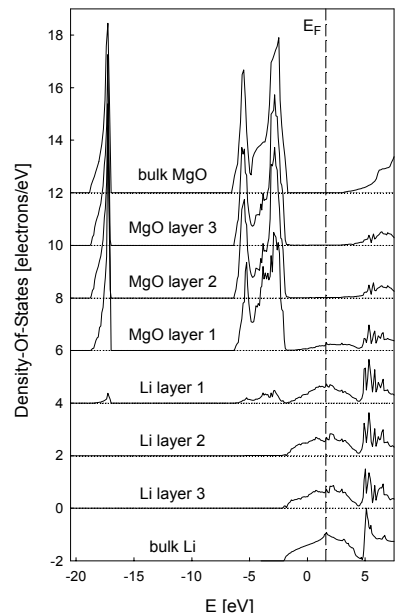

**Fig. 1** Schematic positron affinity profiles $A_+$ of embedded and colloidal nanocrystal systems (a) Cu in Fe [8], (b) Li in MgO [10], (c) Au in MgO [15-17], (d) colloidal CdSe [11,14]. Dashed lines denote schematic positron density profiles $|\psi^+|^2$.

**Fig. 2** Layer-projected energy density-of-states in the first three layers on either side of the Li(100)/MgO(100) interface, calculated using the VASP package with a 20-layer supercell.

**3 Surfaces and nanocrystal structure** Positron studies on the four NC systems in Fig. 1 clearly reveal two important effects of the presence of surfaces or interfaces on the global properties of NCs. First, the interior structure of embedded NCs can differ from the bulk structure due to the relative importance of the interfacial energy term [2]. Second, atoms at surfaces or interfaces often rearrange themselves significantly relative to the bulk geometry [25], leading to surface relaxations or the presence of vacancy-related defects that strongly affect their optical properties. Examples of the first effect include ~1 nm Cu and ~2-5 nm Li NCs which adopt the crystal geometry of the bulk, switching between bcc and fcc structures, respectively, to achieve a (semi-)coherent binding at the interface when placed in bcc Fe or rock salt MgO hosts. This was clearly revealed using positron 2D-ACAR via the observation of the Fermi surface (FS) signatures of bcc-Cu by Nagai *et al.* [8] and of fcc-Li by Falub *et al.* [9] in the NCs. Traditional methods such as transmission electron microscopy or XRD were not successful in resolving the crystal structure, primarily due to the small sizes of the NCs involved and the low atomic number of Li. Notably, the composition of the Cu NCs could be resolved more accurately than in comparative atomic probe studies.

Embedded Au NCs in transparent host media derive their application potential from their interesting Mie scattering optical properties [1, 15, 16]. Two-detector Doppler, lifetime and 2D-ACAR methods reveal the presence of vacancies and small vacancy clusters at the Au/MgO boundary [15-17], which have a pronounced effect on Mie scattering. This is related to a clear change in the local density of states at the Fermi level induced by the presence of the vacancy clusters at the interface [16]. In Fig. 2 we show how the electronic structure near a Li/MgO interface is modified from the bulk over small distances from the interface of the order of one unit cell. The layer-resolved electronic density-of-states (DOS) of the fcc-Li(100)/MgO(100) interface is shown, calculated using the VASP package [26]. This underlines the importance of interfaces in determining the NC properties and the ability of positrons as one of only a few methods to resolve (electronic) structure and related properties of NC interfaces.

The surface is furthermore of prime importance in colloidal NCs, with CdSe NC as the prototype. For example, the luminescence of these colloidal NCs strongly depends on the successful passivation of bonds at their surface, which can be tailored for example by using ligand molecules that bind to either Cd or Se atoms. Fig. 3 shows the ratio of 1D-ACAR profiles of two samples with ~3 nm CdSe nanocrystals of a similar size capped either with TOPO/HDA or pyridine ligand molecules, which both attach preferably to Cd-atoms at the NC surface. While TOPO molecules ($[CH_3(CH_2)_7]_3PO$) contain a phosphine end-group which tightly binds to the Cd-sites at the NC surface, pyridine ($C_2H_5N$) molecules are short and form weaker bonds with Cd. In both cases, the coverage of Se-sites with TOP molecules will be either incomplete or absent, except in a solution phase. Clearly, the observed ACAR profiles in Fig. 3 are very similar, with small differences caused by a small difference in average nanocrystal diameter. This shows that the positron hardly detects the ligand molecules. Rather, the positron is trapped in a shell-like surface state at the boundary of the CdSe nanocrystal [14] (Fig. 1(d)) where it is sensitive to changes at the NC surface. Thus, a first unambiguous experimental confirmation was obtained of the outward relaxation of the Se atoms at the NC surface, while the Cd atoms are pulled in. This was predicted recently as a result of first-principles density functional theory (DFT) calculations by Puzder *et al.* [25]. We note here that a local probe such as EXAFS can be used to obtain insight into atomic arrangements of NCs, such as quantitative bond length information of different atomic shells. However, the information obtained via EXAFS clearly is an average over the entire nanocrystal [27].

Preliminary 2D-ACAR studies were furthermore performed to detect electronic interactions between neighbouring NCs on thin films of pyridine-capped CdSe NCs. The interface between neighbouring NCs was manipulated by removal of the pyridine ligand molecules. This is feasible by gentle heating in a vacuum which leads to a reduction of the average distance between nanoparticles from ~7 Å to ~2 Å [3]. The result is a distinct stepwise change of the shape of the ACAR distribution and the extracted S parameter. In this way evolution of the electronic structure can be monitored as the NCs come closer and CdSe like binding develops at the interfaces and the electrons experience partial loss of confinement within the NCs. These results will be reported elsewhere.

This new way of monitoring properties of NC surfaces using positrons is promising. For progress in applications, control over surface processes and surface composition of NCs is of vital importance. Our studies indicate that positrons can detect sub-monolayer oxygen surface coverages, enabling investigation of surface oxidation processes, which often degrade optical properties of semiconductor NCs. Notable in this connection is the recent development of the first ultra-thin solar cells made of a dual set of CdSe and CdTe NCs [4], which show an encouraging 3% efficiency under standardized solar illumination conditions. The thickness of thin film solar cells typically considered (of the order of ~100 to 1000 nm) matches very well the depth profiling range of positron beams (between ~10 nm and ~2 μm) in the sub-micron range. This further adds to the expectation of significant future contributions in this field using positron methods.

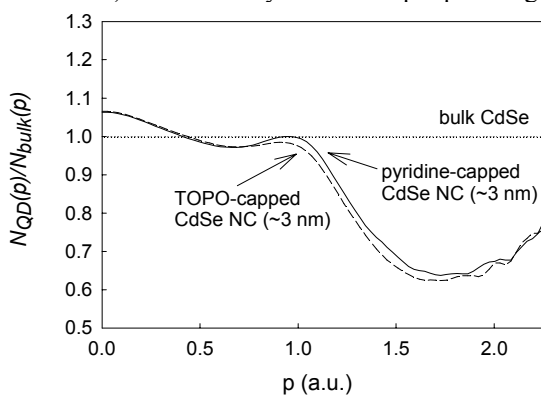
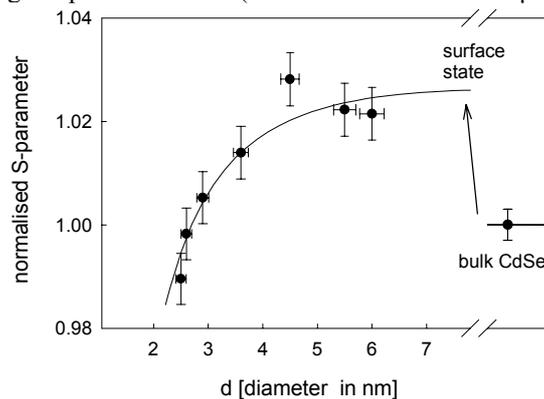

**Fig. 3** Ratio curves of 1D-ACAR profiles $N_{NC}(p) / N_{bulk}(p)$ of pyridine and TOPO-capped CdSe NC thin films, respectively, with reference to the 1D-ACAR curve $N_{bulk}(p)$ of a bulk CdSe single crystal.

**Fig. 4** S-parameter of TOPO-capped CdSe NCs in the size range of ~2 to ~6 nm diameter normalised to the S-parameter of bulk CdSe. The solid line serves as a guide to the eye and represents a $1/d^2$ variation.

**4 Valence electron confinement and electronic structure** One of the most important aspects of NCs is the large variation with size of their electronic structure [2, 3]. Confinement of electron and hole states in semiconductor NCs leads to pronounced changes in the band gap, inducing a strong variation in colour of solutions of semiconductor NCs as a function of their size. Positrons are a very suitable probe for monitoring the electronic structure of NCs via the valence electron momentum density (EMD). The first experimental observation of the size-dependent EMD by Weber *et al.* [11] on colloidal CdSe NCs revealed a clear inverse-square power law of momentum density broadening versus NC diameter, in line with the optical band gap scaling law. Model calculations on the momentum density of the confined homogeneous electron gas (HEG) were presented by Saniz *et al.* [12] who considered confinement in a spherical well potential at various electron densities. These calculations showed a clear broadening at the Fermi momentum in positron annihilation and Compton profiles following a near $1/d$ law with the NC diameter $d$. This behaviour seems to be appropriate for metal NCs. 2D-ACAR studies on bcc-Cu NCs embedded in Fe indeed reveal pronounced broadening at the Fermi momentum with a $1/d$ dependence [28]. Systematic variations in the positron lifetime and shape (S) parameters were predicted from the model calculations. Fig. 4 shows the size variation of the S-parameter for thin films of colloidal CdSe NCs with sizes up to ~6 nm compared to bulk CdSe. The observed reduction is qualitatively similar to the calculated variation [12] and follows a near inverse-square diameter behaviour typical of semiconductor NCs. Future positron studies would benefit from a comparison with first-principles studies of small NCs, which have recently become feasible for sizes up to ~2 nm [25, 29].

**5 Perspectives in positron studies of nanocrystals** New challenges and prospects in this field can be identified. During the past few years, advances in the tailoring of NCs made it possible to assemble them into highly ordered NC superlattices [3]. NCs with rather narrow size distributions are gently deposited to form near hcp or fcc close-packed arrangements. In $e^+$ studies conducted so far, the films of colloidal CdSe NCs were randomly oriented. The anisotropy of a [0002]-projected 2D-ACAR distribution of an oriented CdSe single crystal (Fig. 5), on the other hand, shows clear peaks related to the orientational dependence of its electronic structure. We foresee that this imaging capability of 2D-ACAR can also be applied to resolve the orientation dependence of NC electronic interactions in NC superlattices.

New possibilities also exist in the field of magnetic NCs. First, spin-resolved model calculations [13, 30] show that the momentum density broadening itself is strongly dependent on the magnetic properties of the NCs. Secondly, the spin-resolved electron momentum density of magnetic NCs can be obtained experimentally using radio-isotope $e^+$ sources, which emit partially polarized positrons. This was used by Asoka-Kumar *et al.* [6] to show the non-magnetic behaviour of Cu NCs embedded in ferromagnetic Fe.

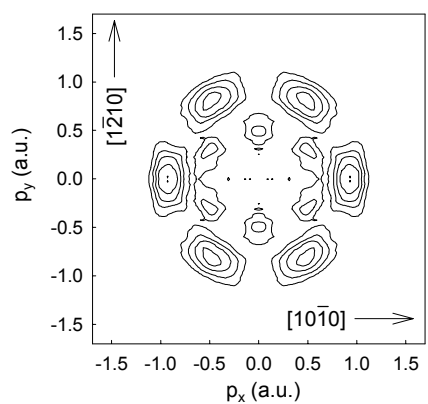

**Fig. 5** Anisotropy of the [0002]-projected 2D-ACAR distribution of a wurtzite CdSe single crystal.

**6  Conclusion**  A short overview of positron studies of nanocrystals is presented, illustrated by recent experimental and theoretical advances. Taking advantage of the positron as a sensitive self-seeking local probe, positron methods are well-suited for investigating structural properties of metallic and semiconductor nanocrystals. These include detection of vacancies at interfaces of embedded nanocrystals, study of the structure and composition of NC surfaces and their (modified) internal crystal structure. Unique insights into the (spin-resolved) electronic structure of colloidal and embedded nanocrystals can be obtained through observation of the momentum density of the valence electrons. Promising areas of application include nanostructured ultra-thin solar cells, embedded metallic NCs for fission and fusion reactor applications, metal-hydride nanocrystals for improved hydrogen storage, as well as semiconductor quantum dots and wires for future opto-electronic and spintronics applications.

**Acknowledgements**   Part of this work is embedded within the Delft Institute for Sustainable Energy (DISE), Delft University of Technology. This work is supported by the U.S. DOE contract DE-AC03-76SF00098, and benefited from the allocation of supercomputer time at NERSC, Northeastern University's Advanced Scientific Computation Center (ASCC) and at SARA, Amsterdam, with support from NCF/NWO (National Computer Facilities of the Netherlands Science Foundation). We thank Dr. Natalia Zaitseva for supply of part of the CdSe NC samples.